# Polarization Decoupling Multi-Port Beam-Splitting Metasurface for Miniaturized Magneto-Optical Trap

Tian Tian[1], Chen Qing[2], Yuxuan Liao[1], Jiajun Zhu[2], Yongzhuo Li[1], Xue Feng[1]*, Dengke Zhang[2]*and Yidong Huang[1]*

[1]Department of Electronic Engineering, Tsinghua University, Beijing 100084, China

[2]School of Instrumentation and Optoelectronic Engineering, Beihang University, Beijing 100191, China

*Corresponding author: x-feng@tsinghua.edu.cn; dkzhang@buaa.edu.cn; yidonghuang@tsinghua.edu.cn

**Abstract**

In regular magneto-optical trap (MOT) systems, the delivery of six circularly polarized (CP) cooling beams requires complex and bulky optical arrangements including waveplates, mirrors, retroreflectors, etc. To address such technique challenges, we have proposed a beam delivery system for miniaturized MOT entirely based on meta-devices. The key component is a novel polarization decoupling multi-port beam-splitting (PD-MPBS) metasurface that relies on both propagation phase and geometric phase. The fabricated samples exhibit high beam-splitting power uniformity (within 4.4%) and polarization purities (91.3%~93.2%). By leveraging such beam-splitting device as well as reflective beam-expanding meta-device, an integrated six-beam delivery system for miniaturized MOT application has been implemented. The experimental results indicate that six expanded beams have been successfully delivered with uniform power (within 9.5%), the desired CP configuration and large overlapping volume (76.2 mm³). We believe that a miniaturized MOT with the proposed beam delivery system is very promising for portable application of cold atom technology in precision measurement, atomic clock, quantum simulation and computing, etc.

## 1. Introduction

Magneto-optical trap (MOT), a cornerstone technique for cooling and trapping neutral atoms, was first successfully demonstrated in 1987 by Raab group at AT&T Bell Laboratory[1,2]. A standard three-dimensional (3D) MOT system consists of a magnetic trap provided by a pair of anti-Helmholtz coils and an optical trap formed by three pairs of mutually orthogonal circularly polarized (CP) laser beams, where the magnetic field zero point coinciding with the center of the optical field[3]. The cold atomic ensembles obtained through MOT would serve as long-coherence-time quantum bits, which are widely applied in the fields of quantum precision measurement[4-7], quantum sensing[8-11], and quantum simulation[12-14]. Specifically, MOT technology facilitates the implementation of atomic optical clocks[15,16], atomic gravimeters[17,18], atomic accelerometers[19], and quantum computing[20,21], establishing it as an indispensable tool in modern atomic physics, geospatial exploration and various foundational scientific research domains.

However, regular MOT systems are typically constrained by their substantial size and weight. One of the primary limitations stems from the required six cooling laser beams with mutually orthogonal propagation directions and specific CP configuration that should be introduced into the vacuum chamber[22]. Generating such multiple beams usually require complex optical systems comprising waveplates, mirrors, retroreflectors, lenses and other components, inevitably leading to bulky system configuration[23-25]. Consequently, the development of chip-based solutions for generating optical fields with desired polarization configurations and propagation directions has emerged as a crucial frontier for MOT related researches. In pursuit of miniaturizing beam delivery system, several innovative MOT configurations have been developed, including pyramidal MOT[26], tetrahedral MOT[27], and grating MOT (G-MOT)[28-31]. Although pyramidal MOT has significantly reduced system size, it is required that the downward input beam undergoes two reflections and traverses the atom cloud to generate the vertically upward beam. This configuration renders the axial beam vulnerable to absorption shielding by the atom cloud and scattering at the pyramid's edges. Alternative configurations, such as tetrahedral and grating MOTs based on the four-beam arrangement, achieve greater compactness due to reduced beam amount. However, these configurations demonstrate compromised performance in terms of atomic trapping capabilities and cooling efficiency when compared to typical six-beam MOT systems.

Recently, metasurface-empowered devices have opened new avenues to miniaturize MOT systems. Metasurface is the two-dimensional array of subwavelength scattering elements, which are referred to as "meta-atoms"[32]. By precisely tailoring the geometric dimensions and spatial orientations of meta-atoms, metasurfaces exhibit overwhelming superiority over traditional optical elements, including novel functionalities such as flexible wavefront shaping[33-35], polarization transformation[36-39], and frequency selectivity[40,41]. These advancements have spurred the development of diverse miniaturized meta-devices, including high-performance metalenses[42-44], vortex beam generators[45], invisibility cloaks[46], and meta-holography[47]. Leveraging the flexible beam shaping and polarization control capabilities of metasurfaces, one can replace bulky optical components with meta-devices to delivery six beams with specific CP configuration and directions required in MOT. Based on metasurfaces, several pioneering researches have demonstrated miniaturized six-beam MOT systems. Among them, PIC-launched structures possess more compact system size, but would suffer from low efficiency due to the coupling of the waveguide mode on chip to propagating mode in free space[48-50], which is quite fatal for quantum applications. For free-space illuminated schemes, despite metasurfaces are employed to partially substitute bulky optical components, the assistance of additional mirrors and retroreflectors remains necessary[51,52], thereby limiting the integration of MOT systems. So far, there is a lack of fully integrated metasurface-based MOT scheme under free-space illuminating.

In this paper, we have proposed and demonstrated a six-beam delivery system for miniaturized MOT. In our work, six large-diameter beams are delivered entirely through meta-devices instead of relying on external bulky optical elements. Specifically, a novel polarization decoupling multi-port beam-splitting (PD-MPBS) based on metasurface is implemented to achieve the functionalities of multiple beam splitting and independent polarization manipulation. For PD-MPBS metasurface, both propagation phase and geometric phase are employed to decouple the amplitude and phase modulation for orthogonal polarization components. For the fabricated three-port PD-MPBS samples, the measured power differences between the splitting sub-beams are within 4.4%, and the polarization purities of the sub-beams are 91.3%~93.2%. Besides, a reflective beam-expanding metasurface is also demonstrated to expand the sub-beams to millimeter-level diameter. By leveraging the aforementioned two kinds of meta-devices, a six-beam delivery system for miniaturized MOT application has been constructed. In the experiment, the generation and overlapping of six expanded beams have been observed. Specifically, the power of each sub-beam surpasses 1.14mW, and the power uniformity across the six beams is maintained within 9.5%. The circular polarization purities of six beams range from 88.5% to 90.6%. Additionally, owing to the mode field diameter (MFD) exceeding 5 mm for each beam, the overlapping volume of the six intersecting beams reaches approximately 76.2 mm³. In summary, the integrated optical system could deliver six beams characterized by uniform power distribution, desired CP configuration, and a large overlapping volume, which are highly promising for MOT systems with enhanced atom capture capacity and ultralow cooling temperature. We anticipate that this metasurface-endowed miniaturized MOT system could serve as a robust platform for portable cold-atom technologies in precision measurement, atomic clock, quantum simulation and quantum computing, etc.

## 2. Results and Discussion
### 2.1. Design Principles

In a standard 3D MOT system (illustrated in **Figure 1a)**, three pairs of mutually orthogonal CP laser beams with equal optical power are required to suppress the random thermal motion of atoms. Among them, two beams are aligned with the magnetic field coil axis (denoted as axial beams) and supposed to possess opposite helicity relative to the other four radial beams. Moreover, the MFDs of all beams are preferable to be millimeter scale so that enough quantity of atoms can be trapped. To reduce the MOT system size and meet the aforementioned requirements simultaneously, we have proposed a beam delivery system for miniaturized MOT based on meta-devices, which is schematically shown in Figure 1b. Such scheme involves two PD-MPBS metasurfaces (each denotated as MS-PBS) and six reflective beam-expanding metasurfaces (denoted as MS-REH). The PD-MPBS metasurface is engineered to deliver three pairs of counter-propagating beams with desired propagation directions and polarization configuration, while the reflective beam-expanding metasurface is designed to enlarge the beam diameter to millimeter scale. In our scheme, to maintain the orthogonality of three pairs of beams, a hexagonal prism glass chamber is also employed, and all metasurfaces are adhered on the corresponding outside surfaces. The chamber was customized to be filled with rubidium (Rb) atomic vapor and the specific size is provided in Figure S1 of Supporting Information.

The schematic of the whole system from lateral view is shown in Figure 1c. For the sake of versatility, we have utilized rectangle nanofins as meta-atoms for both two kinds of meta-devices (MS-PBS & MS-REH) in design. Specifically, for the operation wavelength of 780 nm, which corresponds to the $D_2$ line transition of Rb atoms, amorphous silicon (α-Si) and gold (Au) nanofins are adopted for high transmittance and reflectance, respectively. For the PD-MPBS metasurface, composite phase modulation is employed to achieve three-port polarization-decoupling beam splitting, which will be elaborated in subsequent section. For the reflective metasurface, pure geometric phase is employed, which only depends on the rotation angles of the anisotropic nanofins (See the inset of MS-REH in Figure 1c). Since geometric phase only takes effect on the cross-polarization channels, MS-REH is specifically designed as a half-wave plate (HWP) to avoid the unused co-polarization component. Hence, this meta-device simultaneously realizes three functionalities of reflection, beam-expansion and HWP (hence, abbreviated as REH). Although each REH metasurface introduces additional polarization conversion, chiral transformations would be imposed on all sub-beams. Thus, such arrangement of the beam polarization states can still meet the requirements of MOT system.

In Figure 1c, the beam trajectories are depicted with two colors representing different CP states. In this work, the mode profiles of all beams are specified as fundamental Gaussian mode since it is commonly utilized in spatial optical systems. From the top side of the hexagonal prism cell, an incident right circular polarization (RCP) Gaussian beam is split by the PD-MPBS metasurface (MS-PBS) into one RCP beam and two left circular polarization (LCP) beams with uniform power ratio, and then the sub-beams are directed toward three non-adjacent sidewalls of the hexagonal prism cell. To achieve mutual orthogonality among the splitting sub-beams, the splitting angle of the PD-MPBS metasurface is designed to be 54.7°, corresponding to the included angle of 35.3° with respect to the metasurface plane. Following this, such three sub-beams are individually reflected, expanded and polarization-converted by the corresponding REH metasurfaces (MS-REHs), and finally converge at the center of the cell with expanding MFDs of approximately 5mm. Besides, to generate the other three counter-propagating beams, an identical optical path is symmetrically arranged at the bottom side of the cell. Finally, the overlapped six sub-beams would enable cooling and trapping of the released Rb atomic vapor.

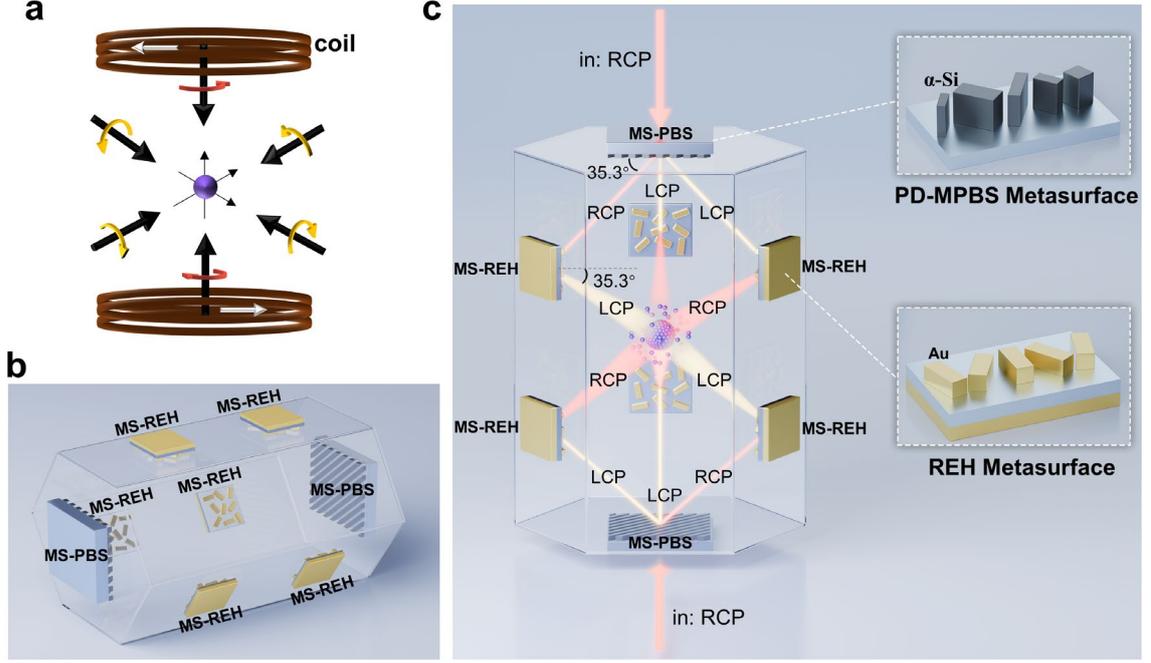

**Figure 1.** The required configuration of cooling laser beams in a standard 3D MOT and the schematic diagrams of the proposed metasurface-based six-beam delivery system for miniaturized MOT. (a) The sketch of the required configuration of cooling laser beams in a standard 3D MOT. (b) The overall schematic of the proposed metasurface-based six-beam delivery system for miniaturized MOT. (c) The lateral view of the metasurface-based beam delivery system and the core meta-devices.

As mentioned above, in our scheme, two kinds of meta-devices are both constructed with anisotropic nanofins. Hence, for generality, Jones matrix formulation is employed for the device design. For a polarization dependent meta-atom, when both the propagation phase and geometric phase are considered, the Jones matrix in CP basis is expressed as:[53]

$$J = R_c(-\theta)\begin{bmatrix} A_{RR}e^{i\phi_{RR}} & A_{RL}e^{i\phi_{RL}} \\ A_{LL}e^{i\phi_{LR}} & A_{RL}e^{i\phi_{LL}} \end{bmatrix} R_c(\theta) = \begin{bmatrix} A_{RR}e^{i\phi_{RR}} & A_{RL}e^{i(2\theta+\phi_{RL})} \\ A_{LR}e^{i(-2\theta+\phi_{LR})} & A_{LL}e^{i\phi_{LL}} \end{bmatrix} \quad (1)$$

where, $\theta$ refers the in-plane rotation angle of the meta-atom, and $R_c(\theta)$ is the rotation matrix. $A_{RR}, A_{RL}, A_{LR}, A_{LL}$ denotes the normalized transmitted amplitude, and the first/second letter in subscript ($R/L$) refers the CP state of the incident/transmitted light, respectively. Similarly, $\phi_{RR}, \phi_{RL}, \phi_{LR}, \phi_{LL}$ denotes the propagation phase.

Additionally, when the nanofin structure is considered as meta-atom, the following relations can be derived based on mirror symmetry: $A_{RR} = A_{LL}, A_{RL}=A_{LR}, \phi_{RR} = \phi_{LL}, \phi_{RL}=\phi_{LR}$. Under these conditions, if the incident light is RCP state, the output electric field $E_{\text{out}}$ can be written as:

$$E_{\text{out}} = JE_{\text{in}} = \begin{bmatrix} A_{RR}e^{i\phi_{RR}} & A_{RL}e^{i(2\theta+\phi_{RL})} \\ A_{RL}e^{i(-2\theta+\phi_{RL})} & A_{RR}e^{i\phi_{RR}} \end{bmatrix}\begin{bmatrix}1\\0\end{bmatrix} = \begin{bmatrix} A_{RR}e^{i\phi_{RR}} \\ A_{RL}e^{i(-2\theta+\phi_{RL})} \end{bmatrix} \quad (2)$$

Further, Equation 2 can be divided into two cases according to the required meta-devices. Firstly, for the REH metasurface, the simplest geometric phase-only modulation is implemented, indicating that the propagation phases ($\phi_{RR}$ and $\phi_{RL}$) equal to 0. Besides, since the nanofins act as local HWPs, Equation 2 can be simplified as $E_{out}^{MS_2} = \begin{bmatrix} 0 \\ e^{-2\theta i} \end{bmatrix}$. Thus, arbitrary phase distribution can be conveniently obtained by simply altering the rotation angles of nanofins via geometric phase principle. On the other hand, for the PD-MPBS metasurface, the geometric phase is incapable of achieving polarization-decoupling. Here, both the propagation phase and geometric phase are employed to increase the design degrees of freedom. As shown in Equation 2, the phase term of the two output polarization components can be manipulated independently. Specifically, the output co-polarization component (i.e., RCP) is modulated solely by propagation phase while the cross-polarization component (LCP) is modulated by both propagation phase and the geometric phase. Additionally, the amplitude coefficient ratio ($A_{RR}/A_{RL}$) can be tailored by selecting the proper meta-atoms. Thus, through combining propagation and geometric phase, it is possible to modulate both the amplitude and phase of the two polarization components independently.

Based on above analyses, different strategies of manipulating the phase response are employed for the design of two desired meta-devices. Subsequently, the specific device structure could be obtained with corresponding strategies. For both meta-devices, the design process can be summarized as three steps. The first step is to design the corresponding target phase function to implement PD-MPBS/REH functionality. Following this, the next step is establishing a proper "meta-atom library" which covers phase modulation range of 0~2π. Finally, by referring to meta-atom library, the target phase function is mapped to the specific distributions of meta-atom structural parameters. In followed sections, the details of designing each meta-device would be elaborated in turn.

Firstly, for the PD-MPBS metasurface, the target phase functions for co- and cross-polarization components are different. For co-polarization (RCP) component, the target phase function (denoted as $\phi_1$) is designed to achieve beam deflecting with deflection angle of 54.7°, corresponding to an ordinary blazed grating profile. While for cross-polarization (LCP) component, the target phase function (denoted as $\phi_2$) is engineered to achieve two-port beam splitting with both deflection angles of 54.7°. In terms of $\phi_2$, to obtain beam-splitting phase pattern with high efficiency and fidelity, an optimization algorithm based on gradient descent is employed. The optimization results are presented in Supporting Information S2 and more detailed algorithm principle can be acquired in our previous work[54]. Secondly, an ingenious meta-atom library is then established through carefully selecting the parameters. The meta-atom structure is considered as amorphous silicon (α-Si) nanofin array on quartz substrate to construct all-dielectric device (**Figure 2a**). For more accuracy, we have employed spectroscopic ellipsometry to obtain the refractive index and absorption coefficient of α-Si material (more details are provided in Figure S3 of Supporting Information). The height of nanofins is set as 500 nm, and the period is $p_x=p_y=340$ nm. Through finite-difference time domain (FDTD) simulation, the transmitted amplitude and phase delay are calculated within length ($L$) and width ($W$) ranging from 60 nm to 300 nm. Figure 2b reveals that the normalized amplitude coefficients are relevant to the specific nanofin sizes and satisfy the complementary relation of $A_{RR}^2 + A_{RL}^2 = 1$. Furthermore, the polarization conversion efficiency (PCE), which is defined as $PCE = A_{RL}^2/(A_{RR}^2 + A_{RL}^2)$, is calculated and plotted in Figure 2c. To achieve three-port beam splitting with uniform power ratio, nanofins with $PCE \approx 2/3$ are carefully selected and marked with black circles in Figure 2c, thereby establishing a meta-atom library with 143 nanofins to satisfy the target power ratio. Meanwhile, for the selected nanofins, the phase delay $\phi_{RR}$ can nearly cover from 0 to $2\pi$ (Figure 2d), which allows for arbitrary target phase function mapping.

Following the established meta-atom library, the target phase function can be independently implemented for both RCP and LCP output components. According to Equation 2, for RCP component, we discretize the target phase function $\phi_1$ into $\psi_1$, which involves 143 phase values. Then, the size distribution ($W$ and $L$) of the nanofins can be obtained by setting $\psi_1 = \phi_{RR}$. Next, through setting $\phi_2 = -2\theta + \phi_{RL}$ (i.e., $\theta = (\phi_{RL} - \phi_2)/2$), the rotation angle distribution of nanofins can be further obtained. Since the selected meta-atoms inherently satisfy that output power of LCP is twice than that of the RCP component, the designed PD-MPBS metasurface can split the incident RCP light into one RCP beam and two LCP beams with equal power ratio. Actually, the finally constructed metasurface is comprised of nanofins with 143 discretized sizes and continuously varying rotation angles.

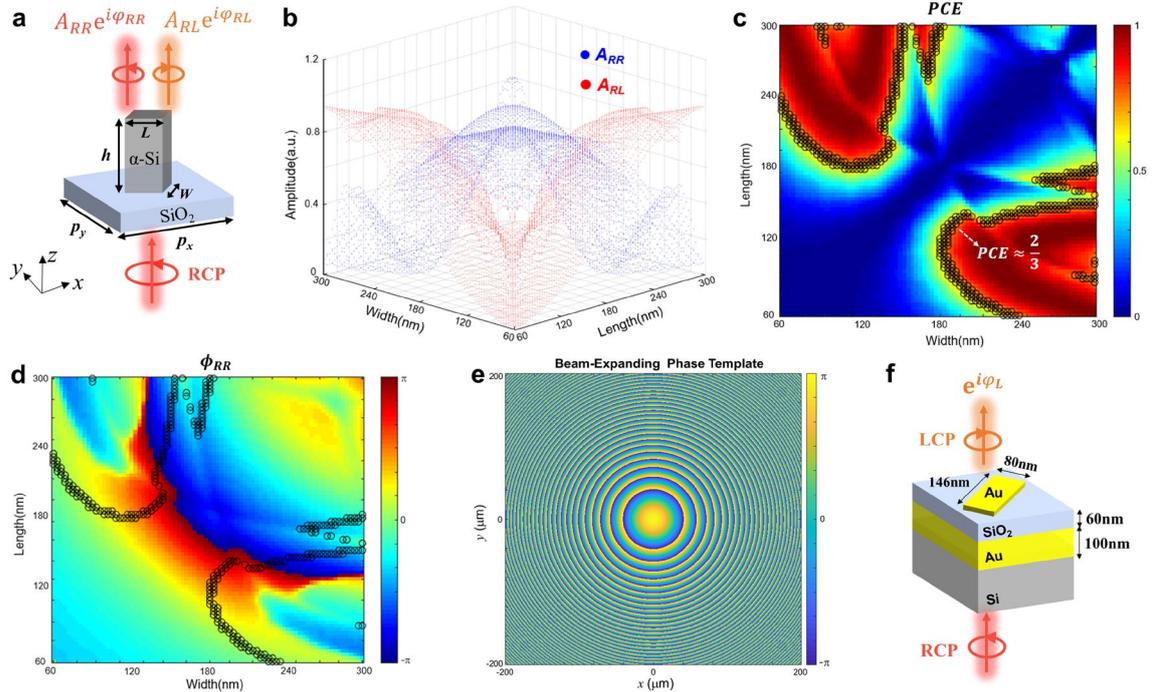

**Figure 2.** Design principles of the proposed PD-MPBS and REH metasurfaces. (a) The schematic of rectangle α-Si nanofin. (b) The normalized amplitude of transmitted RCP and LCP components for different sizes of α-Si nanofins. (c) *PCE* for different sizes of α-Si nanofins. (d) The phase delay of transmitted RCP component for different sizes of α-Si nanofins. (e) The beam-expanding phase template ($f = 1.1$mm) for REH metasurface. (f) The schematic of Au-SiO$_2$-Au nanofin.

Besides the PD-MPBS metasurface, the other required meta-device is the REH metasurface. Similar to the previous design steps, the first aim is to obtain a beam-expanding phase function to enlarge the reflected beam to MFD≈5 mm. The typical beam-expanding phase function is expressed as:

$$\phi_{expanding} = -\frac{2\pi}{\lambda}\left(\sqrt{(x^2+y^2)+f^2}+f\right) \quad (3)$$

where, $\lambda$ is the operation wavelength. $x$ and $y$ are the position coordinates of a specific point on the transverse plane. $f$ refers the focal length of the diverging lens.

To achieve beam expansion to MFD≈5 mm after propagating 16.7 mm (corresponding to the size of the customized hexagonal prism, see details in Supporting Information S1), the focal length ($f$) is iteratively optimized and finally determined to be 1.1 mm. The visualized pattern of the beam-expanding phase template is shown in Figure 2e. Subsequently, meta-atom library of reflective nano-HWPs is also established. The metal-insulator-metal (MIM) structure is employed as meta-atom (see Figure 2f) for high reflectance. The top layer is 20 nm-thick Au rectangular nanofin, and the insulator and bottom layer are 60 nm-thick silicon dioxide ($SiO_2$) and 100 nm-thick Au, respectively. The period is set as $p_x=p_y=200$ nm. Through parameter scanning, the Au nanofin with dimensions of $L$=146 nm and $W$=80 nm is picked out, which can effectively serve as a nano-HWP. After that, the target beam-expanding phase template can be mapped into the rotation angle distribution of Au nanofin array, thereby implementing the desired REH metasurface.

**2.2. Characterization Results of the Proposed Metasurfaces**

To verify our proposal, both the PD-MPBS and REH metasurfaces are fabricated and characterized, respectively. For the fabrication of PD-MPBS metasurface, a 500 nm-thick α-Si layer was firstly deposited on quartz substrate via plasma enhanced chemical vapor deposition (PECVD), and then α-Si nanofin array was prepared via electron beam lithography (EBL) and inductively coupled plasma reactive ion etching (ICP-RIE). The detailed preparation process can be found in Supporting Information S4.1. For the sake of MOT requirements, two samples of PD-MPBS metasurface were fabricated with the same dimension of 200 μm×200 μm. The scanning electron microscope (SEM) images of one sample are presented in **Figures 3a** and 3b. It can be seen that the fabricated α-Si nanofins have distinct dimensions and orientations. On the other hand, for the MIM-type REH metasurface, the Au and $SiO_2$ film was deposited on crystalline silicon (Si) substrate by electron beam evaporation (EBE) and magnetron sputtering in sequence. During the fabrication process, a 5 nm-thick chromium (Cr) adhesion layer was incorporated between each Au layer and Si/$SiO_2$ dielectric medium to prevent the peeling of Au layer. After that, Au nanofins were patterned through EBL and lift-off processes. The diagram of process flow is provided in Supporting Information S4.2. Six samples of REH metasurface were fabricated with the same dimension of 400 μm×400 μm. Figures 3c and 3d present the optical microscope and SEM images of one fabricated sample, respectively. The overall pattern of the sample (Figure 3c) is consistent with the designed phase template (Figure 2e). As shown in Figure 3d, all Au rectangular nanofins possess nearly identical dimensions but varied rotation angles. Additionally, it is worth mentioning that the rectangular edges for both α-Si and Au nanofins have rounded corners. It is attributed to the shorter dwell time of the electron beam, which usually occurs in EBL process.

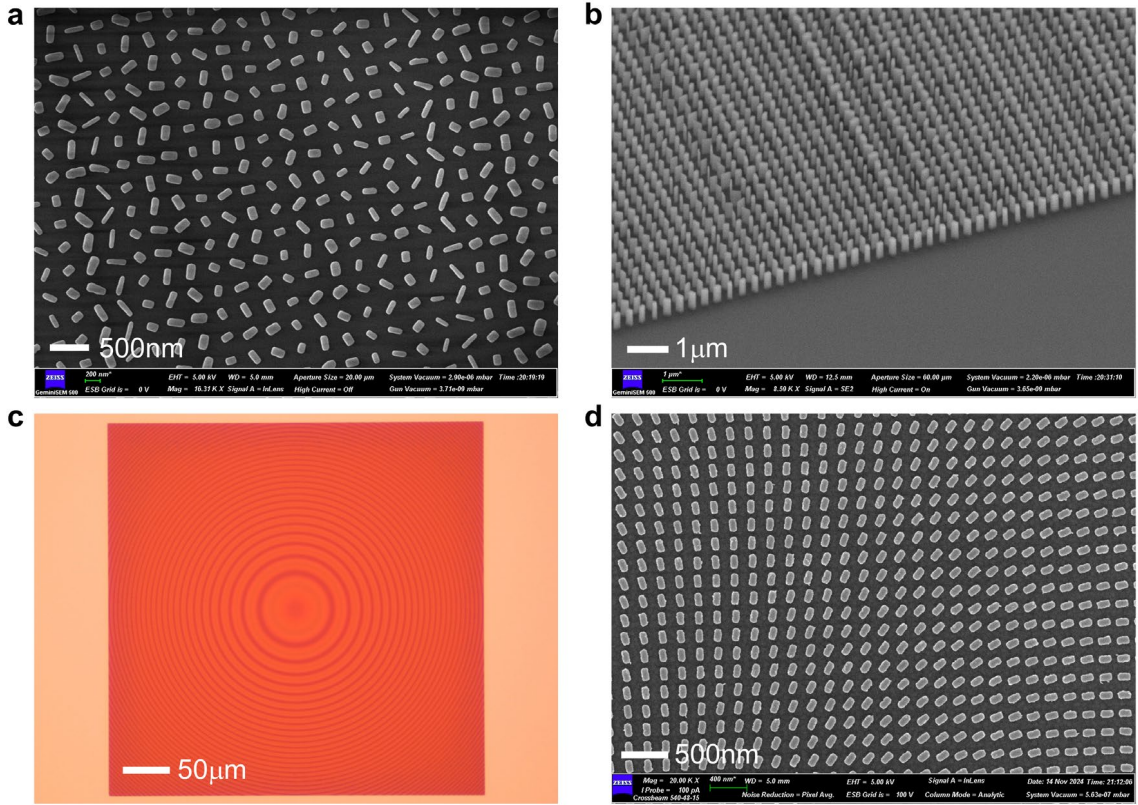

**Figure 3.** The SEM and optical microscope images of the fabricated PD-MPBS and REH metasurface samples. (a) The SEM image of the PD-MPBS metasurface. (b) The SEM image from oblique viewing angle of the PD-MPBS metasurface. (c) The overall optical microscope image of the REH metasurface. (d) The partial SEM image of the REH metasurface.

To characterize the fabricated PD-MPBS samples, a testing system has been built up as schematically shown in **Figure 4a**. The incident laser beam with wavelength of 780 nm is converted into RCP state through multiple wave plates, and subsequently focused onto the metasurface by a lens of $f$=50 mm. The transmitted sub-beams are presented on the observation plane. For each sub-beam, with the recorded distance between the metasurface and observation plane ($\Delta x$), as well as the displacements $\Delta y$ and $\Delta z$ along the $y$-axis and $z$-axis on the observation plane, we can calculate the splitting/deflecting angle $\theta_i$($i$=1,2,3) through:

$$\theta_i = \tan^{-1}\left(\frac{\sqrt{\Delta y_i^2 + \Delta z_i^2}}{\Delta x_i}\right) \quad (4)$$

Additionally, the optical power of each sub-beam is measured by free-space optical power meter. In order to evaluate the polarization state of each transmitted sub-beam, a quarter-wave plate (QWP) followed by a linear polarizer (LP) are settled along the propagation axis of each sub-beam, allowing for recording power of RCP and LCP components independently. Here, we employ the parameter of polarization purity to quantize polarization decoupling efficiency, which is formulated as:

$$Purity(RCP) = \frac{P_R}{P_R+P_L}; \quad Purity(LCP) = \frac{P_L}{P_R+P_L} \qquad (5)$$

where, $P_{R/L}$ refers the measured optical power of RCP/LCP component, respectively.

The measured results for the PD-MPBS metasurface are plotted in Figure 4b. From the original photograph (the inset in the bottom right of Figure 4b), it can be seen that the incident beam is split into three sub-beams after passing through the metasurface. According to the recorded data, three sub-beams are located along a circle with angular separation of ~120° between adjacent two beams, and the beam-splitting angles are calculated to be 54.6°, 54.4°, 54.5° respectively. These results are consistent with the design. It is worth mentioning that the beam-splitting angle (54.7°) corresponds to the phase gradient of 6.6×10$^6$ rad/m, which is quite challenging to achieve for conventional optical devices. To our knowledge, such large phase gradient variation can only be demonstrated with the subwavelength metasurface so far. Additionally, the measured polarization purity for respective sub-beam is 93.2%(RCP), 91.3%(LCP), 92.8%(LCP). For both of the fabricated samples, the power deviations between the three sub-beams are within 4.4% (See complete results of two samples in Table S1 of Supporting Information S5), indicating high beam-splitting ratio fidelity for the proposed PD-MPBS meta-device. From Figure 4b, it can also be noticed that there is a 0-order central spot (unmodulated beam), which accounts for approximately 33.1% of the total transmitted light. The presence of central spot is primarily attributed to fabrication imperfections of nanofins, and it is usually observed in metasurface-related experiments due to the inevitability of fabrication error[51,54,55]. With improved fabrication technics, it is potential to reduce the proportion of central spot and thus increase the diffraction efficiency of metasurface.

On the other hand, the performance of REH metasurface is also measured. As shown in Figure 4c, when the incident Gaussian beam (MFD≈320 μm) is misaligned with the metasurface structure, the light beam reflected by the bottom Au layer of the un-patterned region on the chip would not exhibit beam expansion. In contrast, when properly aligned with the metasurface structure, the reflected light beam would be significantly expanded. The measurement result indicates that the reflected beam could expand to MFD of approximately 5.1 mm after propagating 16.7 mm, which achieves 4.6-fold beam expansion compared to directly reflected light. Besides, in the case of RCP incidence, the polarization purity (LCP) of the output light was measured to be 96.4%, and the average modulation efficiency (i.e., the ratio of reflected power to the incident power) of the fabricated six samples is calculated as high as 88.7%. These experimental results confirm that the REH metasurface performs in accordance with the design.

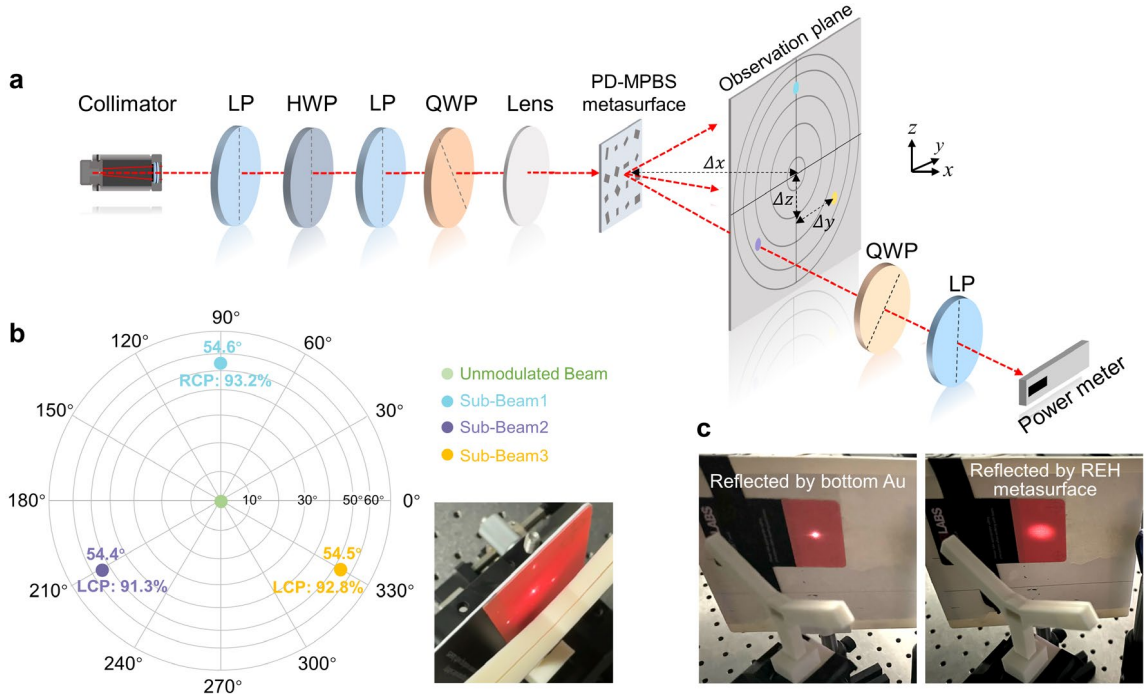

**Figure 4.** The experimental measurement and characterization results of the fabricated PD-MPBS and REH metasurfaces. (a) The schematic of the experimental measurement platform for the PD-MPBS metasurface. (b) Experimental results of the PD-MPBS metasurface. (c) Experimental results of the REH metasurface.

## 2.3. Experimental Setups of Six-Beam Delivery System for Miniaturized MOT

After validating the functionalities of the proposed PD-MPBS and REH metasurfaces, a six-beam delivery system for miniaturized MOT application could be further assembled. Due to the spatial symmetry of the proposed metasurface-based MOT system (see Figure 1c), we firstly implemented half of the optical path to demonstrate the beam property more clearly. A lateral photograph of the half of system setup is shown in **Figure 5a**, which consists of one PD-MPBS metasurface sample (MS-PBS) and three REH metasurface samples (MS-REHs). All the metasurfaces are mounted on customized 3D-printed holders and secured on micro-adjustable stages for fine alignment. The MS-REH samples are positioned at non-adjacent sides of a hexagon base. Additionally, to block the unmodulated 0-order light spot from MS-PBS, an aperture stop is placed along the axis of the light path. Owing to the sufficiently large splitting angle (54.7°), the aperture would not block the desired splitting sub-beams. Through meticulous alignment, the incident 780 nm laser beam can be split by MS-PBS, and then reflected, expanded and polarization-converted by corresponding MS-REHs. Finally, three expanded sub-beams would overlap in the central region along the axial direction (Figure 5b). From Figure 5b, it can be observed that the overlapping region exhibits a "quasi-hexagonal star" distribution. This phenomenon is due to the fact that the detection plane is tilted at an angle of 54.7° relative to the propagation direction of the beams, resulting in an axial elongation effect of each beam. It is important to note that when the detection plane is perpendicular to the propagation direction of each beam, the expanded beams would maintain the Gaussian mode field distribution, with detailed illustrations provided in Figures S6 of Supporting Information.

Furthermore, by employing the same method to implement the other half of optical setup, we have built up a complete six-beam delivery system for miniaturized MOT, as shown in Figure 5c. The system is left-right symmetric, and the portions enclosed by the yellow dashed line both correspond to Figure 5a. The overall size of the metasurface assembly region is about 9.8 mm×5.2 mm×5.2 mm, which exhibits dramatical footprint reduction compared to traditional MOT system. In the experiment, the 780 nm laser source is uniformly divided into two paths by a 1×2 fiber coupler, which are collimated as free-space beams from left and right sides separately. Subsequently, each incident beam is converted to RCP state and focused onto MS-PBS by a series of wave plates and lens. Through the modulation of metasurface assembly, six expanded beams with specific CP configuration would overlap at the center of the system. **Table 1** summarizes the power, MFDs and polarization purities of the six sub-beams. It can be noticed that the power of sub-beams 1~3 are slightly higher than that of sub-beams 4~6. This is due to the splitting ratio deviation of the 1×2 fiber coupler, which results in unequal incident power for two sides of the system (16.9 mW vs. 16 mW). Nonetheless, the power differences among the six beams are within 9.5%. Besides, the polarization purities of all six beams exceed 88.5%. It is worth mentioning that the polarization purities of sub-beams in the assembled system is slightly deteriorated with that measured in individual MS-PBS (Figure 4b). It is the result of the superposition of polarization conversion efficiency introduced by the MS-REH. Additionally, we also simulated and evaluated the beam overlap volume, which is another key indicator of MOT system. Based on the measured characteristics of all sub-beams, the schematic diagram of overlapping scenario obtained by simulation is illustrated in Figure 5d. Since all the expanded beams are Guassian modes, the scenario is similar to the intersecting of three mutually perpendicular cylinders, and the overlap region forms a 3D Steinmetz solid (see the inset in Figure 5d). Thus, the beam overlap volume is ~76.2 mm$^3$, which can be approximately estimated by:

$$V = (2 - \sqrt{2})d_1 d_2 d_3 \qquad (6)$$

where $d_1, d_2, d_3$ denotes the MFDs of the three mutually perpendicular sub-beams delivered from one side of the system.

In summary, it can be concluded that the demonstrated system can deliver six beams with nearly uniform power, MFD exceeding 5 mm, the desired CP configuration as well as overlapping volume of 76.2 mm$^3$. For subsequent research, the transparent hexagonal prism vapor chamber of Rb atoms can be settled inside the metasurface assembly region to demonstrate atom cooling and trapping.

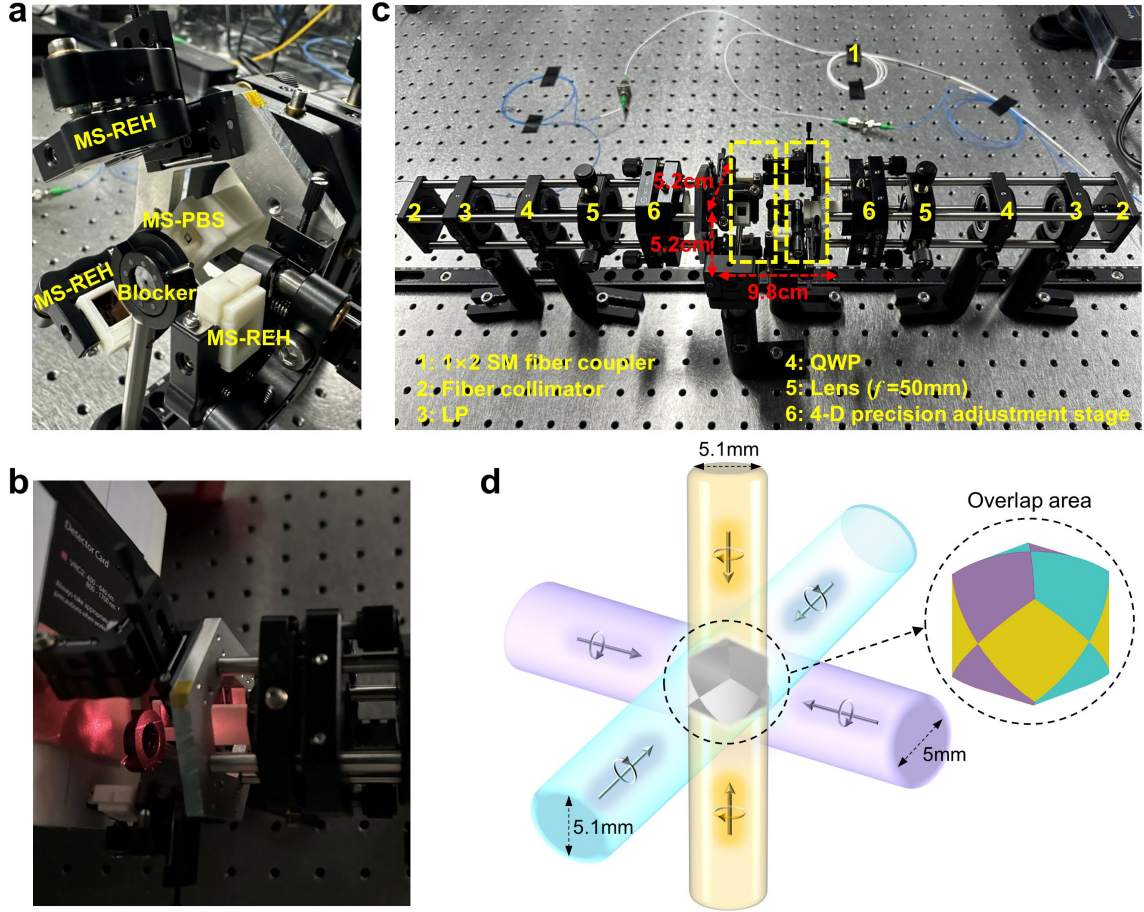

**Figure 5.** The experimental setup and characterization of metasurface-based beam delivery system for miniaturized MOT. (a) The lateral photograph of half of the delivery system. (b) The overlapping photograph of three expanded beams for half of the delivery system. (c) The complete six-beam delivery system based on metasurfaces for miniaturized MOT. (d) The overlapping scenario of six expanded sub-beams in simulation.

**Table 1.** The characterization results of the delivered six sub-beams.

|            | Power(mW) | MFD(mm) | Polarization Purity |
|------------|-----------|---------|---------------------|
| **Sub-beam 1** | 1.26 | 5.1 | 90.0%(LCP) |
| **Sub-beam 2** | 1.22 | 5.0 | 88.5%(RCP) |
| **Sub-beam 3** | 1.20 | 5.1 | 90.3%(RCP) |
| **Sub-beam 4** | 1.16 | 5.1 | 89.7%(LCP) |
| **Sub-beam 5** | 1.14 | 5.0 | 89.1%(RCP) |
| **Sub-beam 6** | 1.14 | 5.1 | 90.6%(RCP) |

## 3. Conclusion

In summary, to achieve compact MOT systems, we have proposed a miniaturized beam delivery scheme based on novel meta-devices. Specifically, a PD-MPBS is demonstrated to split single CP incident beam into multiple RCP and LCP beams with decoupling power ratio and specific polarization configuration. For the fabricated samples of PD-MPBS, the measured power differences between the splitting sub-beams are within 4.4%, and the polarization purities of the sub-beams are 91.3%~93.2%, along with large splitting angle of 54.7°. Additionally, a REH metasurface is also achieved to expand the sub-beams to the desired diameters. With such two kinds of meta-devices, a fully integrated six-beam delivery system for miniaturized MOT application is implemented. With the laser input power of 32.9 mW, the power of each sub-beam exceeds 1.14 mW and the corresponding emission efficiency is significantly higher than that of previously reported PIC-launched miniaturized MOT schemes[48-50] and also competitive with free-space metasurface-based MOT schemes[51,52]. The power level can be further increased by employing higher-power lasers. Besides, the power differences among the six beams are within 9.5%, and the polarization purities are 88.5%~90.6%. Benefiting from MFD of over 5 mm for each beam, the overlap volume of the six intersecting beams is approximately 76.2 mm³. In brief, the demonstrated system is capable to deliver six beams with uniform power, the desired CP configuration and large overlapping volume, which are promising to achieve a MOT system with considerable number of capturing atoms and low cooling temperature. We believe that such a metasurface-based miniaturized MOT provides an effective solution for portable application of cold atom technology in precision measurement, quantum simulation and large-scale quantum computing etc. Moreover, based on the proposed composite phase design, arbitrary PD-MPBS meta-devices can be constructed for various applications. Hence, the proposed PD-MBPS metasurface is not only applicable to miniaturize MOT system, but also holds broad potential for other scenarios requiring multi-port beam splitting and polarization control, such as polarized light detection and ranging (LiDAR) 3D imaging, augmented reality (AR) displays and laser guidance systems.

**Supporting Information**
Supporting Information is available from the author.

**Acknowledgements**
This research was supported by the National Key Research and Development Program of China (2023YFB2806703), and the National Natural Science Foundation of China (Grant No. U22A6004, 92365210，62175124) is greatly acknowledged. This work was also supported by the project of Tsinghua University-Zhuhai Huafa Industrial Share Company Joint Institute for Architecture Optoelectronic Technologies (JIAOT), Beijing National Research Center for Information Science and Technology (BNRist), Frontier Science Center for Quantum Information, Beijing academy of quantum information science, and Tsinghua University Initiative Scientific Research Program.

## Conflict of Interest

The authors declare no conflict of interest.

## Author Contributions

D.Z., T.T., C.Q. and X.F. conceived the idea. T.T. designed and performed the simulations, experiments, and data analysis. Y.L. (Yuxuan Liao) contributed significantly to the phase pattern optimization algorithm based on gradient descent. J.Z. assisted to characterize the performance of metasurfaces. Y.L. (Yongzhuo Li) provided useful discussions and comments. T.T. and X.F. wrote the paper. D.Z., C.Q. and Y.H. revised the manuscript. The manuscript was written through contributions of all authors. All authors have given approval to the final version of the manuscript.